\def\rfr#1{eq. (\ref{#1})}

\def\dert#1#2{\frac{{{d}}{#1}}{{{d}}{#2}}}              

\def\bar{\begin{eqnarray}}
\def\ear{\end{eqnarray}}
\def\bb{\bibitem}
\def\eqi{\begin{equation}}
\def\eqf{\end{equation}}
\def\eqia{\begin{eqnarray}}
\def\eqfa{\end{eqnarray}}
\def\rp#1#2{{#1\over#2}}

\def\lb#1{\label{#1}}

\def\cyg{Cygnus X-1}
\def\hde{HDE 226868}

\def\oc2{$\mathcal{O}(c^{-2})$}

\documentclass{aastex}
\usepackage{spr-astr-addons}
\usepackage{url}

\RequirePackage{color}

\newcommand{\emaila}{lorenzo.iorio@libero.it}

\begin{document}

\title{On the orbital and physical parameters of the \hde/\cyg\ binary system}
\shorttitle{On the parameters of  HDE 226868/Cygnus X-1}
\shortauthors{L. Iorio}

\author{Lorenzo Iorio\altaffilmark{1} }
\affil{INFN-Sezione di Pisa. Permanent address for correspondence: Viale Unit\`{a} di Italia 68, 70125, Bari (BA), Italy.}

\email{\emaila}

\begin{abstract}
In this paper we explore the consequences of the recent determination of the mass $m=(8.7\pm 0.8)M_{\odot}$ of \cyg, obtained  from the Quasi-Periodic Oscillation (QPO)-photon index correlation scaling, on the orbital and physical properties of the binary system \hde/\cyg.  By using such a result and the latest spectroscopic optical data of the \hde\ supergiant star we get $M=(24\pm 5)M_{\odot}$ for its mass. It turns out that deviations from the third Kepler law significant at more than 1-sigma level would occur if the inclination $i$ of the system's orbital plane to the plane of the sky falls outside the range $\approx 41\div 56$ deg: such deviations cannot be due to the first post-Newtonian (1PN) correction to the orbital period because of its smallness; interpreted in the framework of the Newtonian theory of gravitation as due to the stellar quadrupole mass moment $Q$, they are unphysical because $Q$ would take unreasonably large values. By conservatively assuming that the third Kepler law is an adequate model for the orbital period we obtain $i=(48\pm 7)$ deg which yields for the relative semimajor axis $a=(42\pm 9)R_{\odot}$ ($\approx 0.2$ AU).
\end{abstract}


%
      \section{Introduction}
      \cyg, one of the brightest high-energy sources in the sky, with an average $1-200$
keV energy flux of $\sim3\times 10^{-8}$ erg cm$^{-2}$ s$^{-1}$, was discovered  by \citet{Gia67}   and soon became one of the most intensively studied X-ray sources.
      After the detection of its rapid X-ray variability \citep{Oda71} and the identification of its optical counterpart with the O9.7 Iab supergiant star HDE 226868 \citep{Bol72, Web72}, it was considered as one of the
      most likely black hole candidates. The X-ray emission in
Cygnus X-1 is powered mainly by accretion from the strong stellar wind from \hde\ \citep{Pet78}.

      In this paper we investigate some features of the orbital geometry of the \hde/\cyg\ binary system in order to check the compatibility of the so far obtained spectroscopic data of \hde\ \citep{Gie03} with the tight constraints on the black hole mass $m=(8.7\pm 0.8)M_{\odot}$ recently determined by \citet{Sha07} who exploited the existing correlation \citep{Sha06} between the low frequency Quasi-Periodic Oscillations (LF QPO) and the photon index of the power law spectral component of the \cyg\ X-ray emission. We will look for deviations from the third Kepler law, which will not be assumed a priori valid, by discussing the conditions at which they occur and their physical plausibility.
      \section{Deviations from the third Kepler law}
      The analysis of 115 optical spectra of the H$\alpha$ emission line of \hde\ for the 1998-2002 interval allowed \citet{Gie03} to phenomenologically determine the orbital period $P$ and the projected velocity semiamplitude $K$ (see Table \ref{tavola}): for $P$ the same value, within the errors, of the one obtained by \citet{Bro99}, who used  a 26-yr data set,    was obtained. A previous orbital fit can be found, e.g., in \citep{Gie82}.

\begin{table}[t]
\caption{Relevant orbital parameters of the \hde/\cyg\ binary system. $P$ is the orbital period \citep{Bro99}, $K$ is the projected velocity semiamplitude \citep{Gie03}, $q\equiv m/M$ is the ratio of the black hole to the star mass \citep{Gie03}, $m$ is the black hole's mass \citep{Sha07}. All of them have been determined independently of the third Kepler law: $P,K,q$ from optical spectroscopy of \hde, and $m$ from the analysis of certain properties of the \cyg\ X-ray emission.\label{tavola}}
\begin{tabular}{@{}llll}
\hline
$P$ (d)& $K$ (km s$^{-1}$) & $q$ & $m$ ($M_{\odot}$)\\
\tableline
 $5.599829(16)$ & $75.6(7)$ & $0.36(5)$  & $8.7(8)$\\
\hline
 \end{tabular}
\end{table}
      From $P$ and $K$, in turn, it is possible to obtain the stellar projected barycentric semimajor axis $x_M\equiv a_M\sin i$, where $i$ is the inclination angle of the orbital plane to the plane of the sky, as
      \eqi x_M=\rp{PK}{2\pi}=(8.37\pm 0.07)R_{\odot}.\lb{sma}\eqf

      An important result of the spectroscopical analysis by \citet{Gie03} is that the emission component of the radial velocity is compatible with a mass ratio \eqi q\equiv\rp{m}{M} = 0.36\pm 0.05.\eqf The recent result by \citet{Sha07} for $m$ yields for the mass of \hde
      \eqi M=(24\pm 5)M_{\odot};\eqf it disagrees with $M=17.8M_{\odot}$ by \citet{Her95} which, among other things, relies upon various estimates of $i$. Even larger is the disagreement with $M=(40\pm 5)M_{\odot}$ obtained by \citet{Zio05} with evolutionary calculation for the most likely intervals of the
values of the distance and of the effective temperature\footnote{Extending the intervals of
these parameters to 1.8 to 2.35 kpc and 28000 to 32000
K, \citet{Zio05} obtained a range $29\div 50$ for $M$.}:
1.95 to 2.35 kpc and 30000 to 31000 K. It is important to note that the method by \citet{Sha07} is not only independent of the lingering uncertainty in the system's distance from us, but it was also robustly tested by reproducing the masses of other black hole candidates, like the one in the microquasar GRS $1915+105$ \citep{Cas92},  obtained with IR \citep{Gre01} and X-ray \citep{Shr03} observations.  Thus, in the following we will trustworthy rely upon such a result. Note that the mass estimate by \citet{Sha07} for $m$ is well within the  range $4.8 M_{\odot} < m < 14.7 M_{\odot}$ obtained by  \citet{Her95}.

      We are now in the position of looking for genuine deviations from the third Kepler law by comparing the phenomenologically determined period $P$ to the calculated Keplerian one which, from \rfr{sma} and \eqi a=\left(1+\rp{1}{q}\right)\rp{x_M}{\sin i}\lb{smarel}\eqf for the relative semimajor axis, can be expressed  as
      \eqi P^{\rm Kep}=\left(\rp{1+q}{q}\right)\sqrt{\rp{1}{2\pi Gm}\left(\rp{KP}{\sin i}\right)^3}\lb{pkep}\eqf
      in terms of quantities determined independently of the third Kepler law itself: note that \rfr{pkep} does not depend on $M$.
      The overall uncertainty $\delta P^{\rm kep}$ in \rfr{pkep}, which will be considered as a function of $i$, can be conservatively  evaluated by linearly adding the various mismodelled terms due to $\delta q,  \delta G, \delta m, \delta K, \delta P$ as
      \begin{equation}\begin{array}{lll}
      \left.\delta P^{\rm Kep}\right|_q\leq \rp{P^{\rm Kep}}{(1+q)}\left(\rp{\delta q}{q}\right),\\\\
      \left.\delta P^{\rm Kep}\right|_G\leq  \rp{P^{\rm Kep}}{2}\left(\rp{\delta G}{G}\right),\\\\
      \left.\delta P^{\rm Kep}\right|_m\leq  \rp{P^{\rm Kep}}{2}\left(\rp{\delta m}{m}\right),\\\\
      \left.\delta P^{\rm Kep}\right|_K\leq  \left(\rp{3}{2}\right)P^{\rm Kep}\left(\rp{\delta K}{K}\right),\\\\
      \left.\delta P^{\rm Kep}\right|_P\leq  \left(\rp{3}{2}\right)P^{\rm Kep}\left(\rp{\delta P}{P}\right).
      \lb{erori}
      \end{array}\end{equation}

\begin{figure}[t]
   \includegraphics[width=\columnwidth]{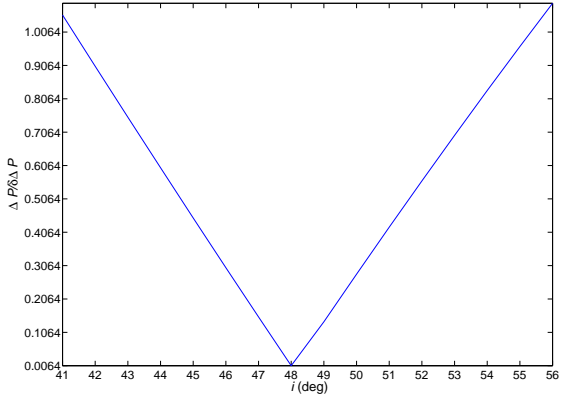}
   \caption{Discrepancy between the phenomenologically measured orbital period $P$ and the computed Keplerian one $P^{\rm Kep}$ as function of the inclination $i$.}
   \label{Figu}
   \end{figure}

   In Figure \ref{Figu} we plot the ratio \eqi \rp{\Delta P}{\delta (\Delta P)}\equiv\rp{|P-P^{\rm Kep}|}{\delta P+ \delta P^{\rm Kep}}\eqf as a function of the inclination;
   outside the range $41$ deg $\lesssim i \lesssim 56$ deg, in which the purely Keplerian period is an adequate model for the measured orbital one, deviations from the third Kepler law significant at more than 1-sigma level would occur; in particular, the estimate $i = (30\pm 7)$ deg by \citet{Gie03} would yield a discrepancy significant at $4/2$-sigma-level, while for $i=35$ deg, chosen by \citet{Her95} because lying in the middle of almost all determinations in literature, the deviation from the third Kepler law would amount to 2-sigma. Can such deviations be considered physically meaningful?

   In principle, also the first post-Newtonian (1PN) correction to the Keplerian period \citep{Dam86},
   \eqi P^{\rm 1PN} = \rp{3}{\sqrt{2}c^2}\left(\rp{1+q}{q}\right)\sqrt{\rp{\pi GmKP}{\sin i}}\left[1-\rp{q}{3(1+q)^2}\right]\lb{Damp}\eqf here expressed in terms of the phenomenologically determined quantities,
   should be taken into account in modeling the orbital period; however, it is not possible that \rfr{Damp} is the cause of  the significant discrepancies because it is six orders of magnitude smaller than $\delta P^{\rm Kep}$ itself for all values of $i$.   Note that considering the possibility that 1PN terms may have some observable effects in so close binary systems should not be considered trivial since for, e.g., the exoplanet HD209458b the accuracy in determining its 3.5 d orbital period would, in fact, allow to detect\footnote{The uncertainties in the system's parameters induce systematic errors which prevent from implementing such a goal.} the relativistic correction \citep{Ior06}.

\begin{figure}[t]
   \includegraphics[width=\columnwidth]{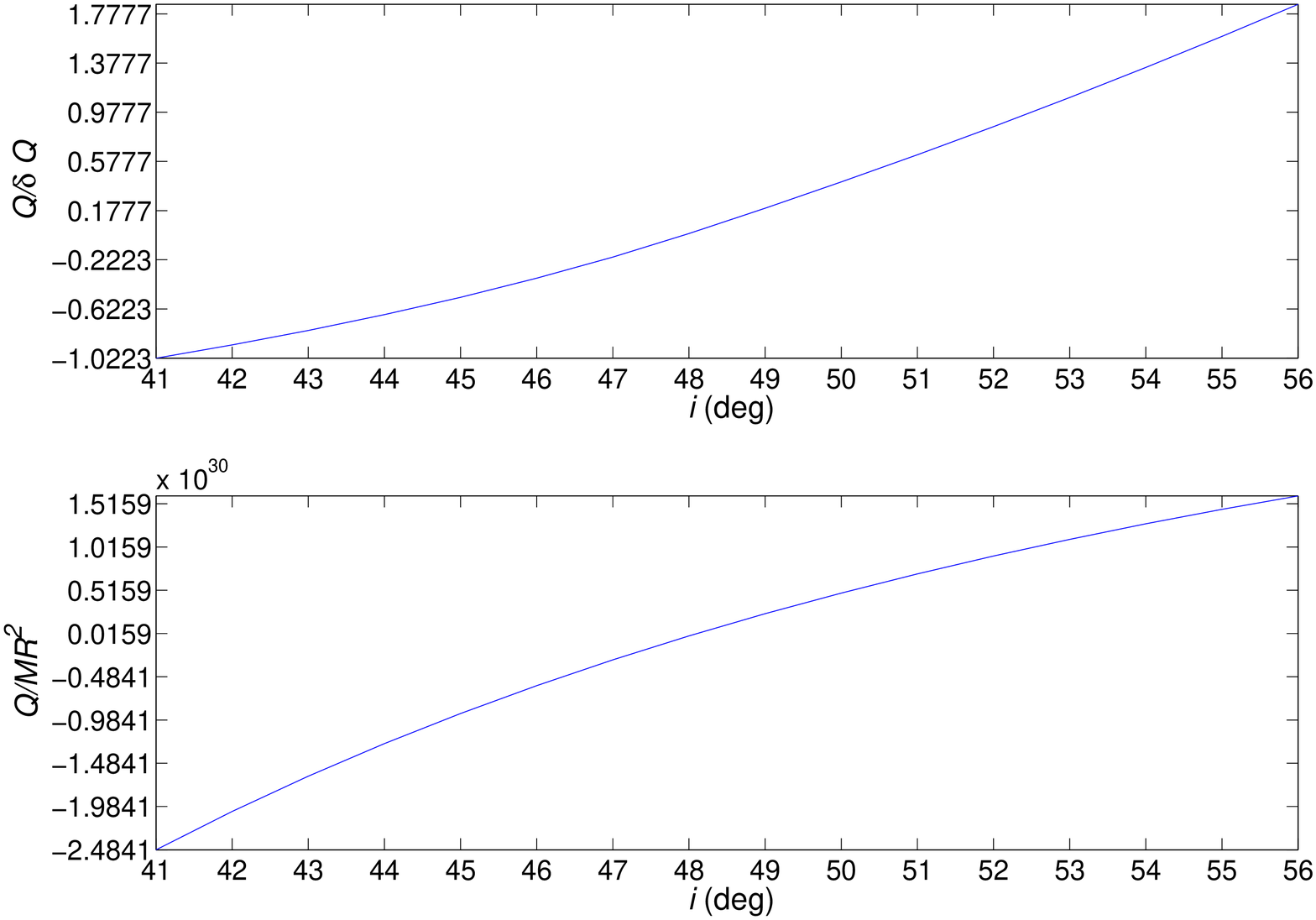}
   \caption{Upper panel: $Q/\delta Q$. Lower panel: $Q/MR^2$ as functions of the inclination $i$.}
   \label{alura}
   \end{figure}

   Let us see what happens if we add to the Keplerian period the correction due to the quadrupole mass moment $Q$ of the system: in terms of the determined parameters we have, from \rfr{QU},
   \eqi Q=\rp{\kappa^2}{3\sqrt{2}}\sqrt{\left(\rp{mP}{\pi}\right)^3\rp{GK}{\sin i}} -\rp{m\kappa^3}{6}\left(\rp{KP}{\pi\sin i}\right)^2\lb{Q},\eqf
   where we have posed
   \eqi \kappa\equiv \rp{1+q}{q}.\eqf   The total uncertainty in $Q$ can be conservatively evaluated by summing the various contributions
    \begin{equation}\begin{array}{lll}
    \left.\delta Q\right|_G\leq  \left|\rp{\kappa^2}{ 6\sqrt{2} }\sqrt{ \left(\rp{mP}{\pi}\right)^3\rp{K}{G\sin i} }\right|\delta G,\\\\
     \left.\delta Q\right|_m\leq   \left|\rp{\kappa^2}{2\sqrt{2}}\sqrt{\left(\rp{P}{\pi}\right)^3\rp{GmK}{\sin i}} -\rp{\kappa^3}{6}\left(\rp{KP}{\pi\sin i}\right)^2\right|\delta m,\\\\
      \left.\delta Q\right|_P\leq   \left|\rp{\kappa^2}{2\sqrt{2}}\sqrt{\left(\rp{m}{\pi}\right)^3\rp{GPK}{\sin i}} -\rp{mP\kappa^3}{3}\left(\rp{K}{\pi\sin i}\right)^2\right|\delta P,\\\\
      \left.\delta Q\right|_K\leq   \left|  \rp{\kappa^2}{ 6\sqrt{2} }\sqrt{ \left(\rp{mP}{\pi}\right)^3\rp{G}{K\sin i} }  -\rp{mK\kappa^3}{3}\left(\rp{P}{\pi\sin i}\right)^2  \right|\delta K,\\\\
       \left.\delta Q\right|_q\leq   \left| \rp{\sqrt{2}}{3}\rp{\kappa}{q^2}\sqrt{\left(\rp{mP}{\pi}\right)^3\rp{GK}{\sin i}}-\rp{m}{2}\left(\rp{\kappa}{q}\right)^2\left(\rp{KP}{\pi\sin i}\right)^2\right|\delta q.
      \lb{erroriq}
      \end{array}\end{equation}
In Figure \ref{alura}
we plot $|Q|/\delta Q$ (upper panel) and\footnote{We used $R=17R_{\odot}$ \citep{Her95}; according to \citet{Zio05}, $R/R_{\odot}= 10.59 d$,  with $d$ in kpc: $d\approx 1.8-2.35$ kpc.} $Q/MR^2$ (lower panel): we see that ascribing the deviation from the third Kepler law, when it becomes significant,  to the \hde's quadrupole mass moment leads to meaningless results since $Q$ would assume values incompatible with zero at more than 1-sigma level, but unreasonably large.

By conservatively assuming that no deviations from the third Kepler law occur, given the present-day level of accuracy in knowing the system's parameters, it is possible to determine $i$ as
\eqi i =\arcsin \left[KP^{1/3}\left(2\pi Gm \kappa^{-2}\right)^{-1/3}\right]=(48\pm 7)\ {\rm deg}.\eqf

In regard to the inclination $i$, \citet{Gie03} obtain \eqi i = (30\pm 7)\ {\rm deg}\lb{loli} \eqf from the measured emission and absorption
      radial velocities  and by making certain assumptions about isotropy of the outflow velocity; after noting that their result is compatible with other estimates\footnote{Optical polarimetry of \hde\ yields $25$ deg $\leq i \leq 67$ deg \citep{Dol89}   and $20$ deg $\leq i \leq 40$ deg \citep{Dan81}. } like $28$ deg $\leq i \leq 38$ deg from light curve analysis by \citet{Gie86a} and $10$ deg $\leq i \leq 40$ deg  from fitting the X-ray emission curve by \citet{Wen99}, \citet{Gie03} conclude that \rfr{loli} is likely a lower limit.
Our result, which is incompatible with the one by \citet{Gie03},  being, instead, compatible with, e.g., the result by    \citet{Dol89}, can be useful in terms of anisotropy in the wind outflow velocity \citep{Fri82}.

\begin{figure}[t]
   \includegraphics[width=\columnwidth]{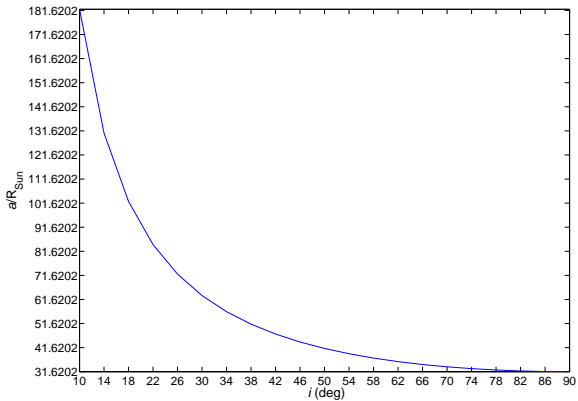}
   \caption{Relative semimajor axis $a$, in Solar radii, as function of the inclination $i$.}
   \label{dista}
   \end{figure}

It is also interesting to evaluate  the relative separation of \hde\ and \cyg\ (see Figure \ref{dista}): such an information may be helpful about \hde's dimensions.
   %
   %
The relative semimajor axis, according to \rfr{sma} and \rfr{smarel} which are independent of the third Kepler law and on the mass $m$ of \cyg,
 amounts to $a=(42\pm 9)R_{\odot}$ ($\approx 0.2$ AU) for $i=48$ deg; the minimum value, i.e. $a=31.6R_{\odot}$ (0.1 AU), would occur for the edge-on configuration while for $i=10$ deg it amounts to about 181 Solar radii (0.8 AU).

\begin{figure}[t]
   \includegraphics[width=\columnwidth]{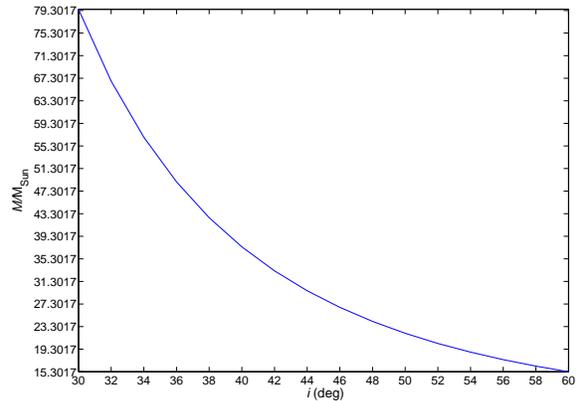}
   \caption{Mass of \hde, in Solar masses, as function of the inclination $i$. The third Kepler law was used.}
   \label{arreit}
   \end{figure}

 Note that, should one consider $M=(40\pm 5)M_{\odot}$  inferred by \citet{Zio05}, the black hole's mass  by \citet{Sha07} would yield $q=0.22\pm 0.05$. Such a value, which is inconsistent at 1.4-sigma level with the one by \citet{Gie03}, would lead to unreasonably large values for the quadrupole mass moment for $i \lesssim 45$ deg, being all the other values up to 90 deg compatible with zero $Q$ and no deviations from the third Kepler law.
 Near edge-on orbital configurations imply that eclipses in the X-ray emission of \cyg\ would be allowed ($i\gtrsim 65$ deg), but they have never been detected.  Thus, the estimate for the \hde's mass by \citet{Zio05} does not seem plausible. Another line of reasoning is the following. From the third Kepler law we can express the \hde's mass in terms of the spectroscopically determined parameters as
 \eqi
 GM = \left(\rp{P}{2\pi}\right)(1+q)^2\left(\rp{K}{q\sin i}\right)^3\lb{massaM};
 \eqf
 as can be noted from Figure \ref{arreit},
$M$ assumes the values by \citet{Zio05}, who relies on the third Kepler law, for 37 deg $\lesssim i \lesssim$ 41 deg; but our previous analysis, independent of $M$ itself, has shown that for such values significant deviations from the third Kepler law would occur (see Figure \ref{Figu}).

Note that our results are compatible with a system distance of about 1.80 kpc \citep{Mal97} and an effective temperature $T_e=32000$ K \citep{Her95} of \hde, according to Table 1 of \citep{Zio05} based on the quite general approach by \citet{Pac74}.

\section{Discussion and conclusions}
In this paper we refined some orbital and physical parameters of  the \hde/\cyg\ binary system    by exploiting  the latest  spectroscopic optical data from \hde\ and the recent determination of the mass $m$ of \cyg\ via QPO-index correlation scaling. For the mass of the supergiant star we obtained $M=(24\pm 5)M_{\odot}$, which is in disagreement with other estimates present in literature. Then, we constrained the inclination $i$ by looking for statistically significant  deviations from the third Kepler law which would occur, at more than 1-sigma level, for $i\lesssim 41$ deg and $i\gtrsim 56$ deg. After noting that they cannot be due to the 1PN correction to the Keplerian period, we concluded that, if attributed to the star's quadrupole mass moment $Q$, they are to be considered unphysical because of the too large values which $Q$ would take in this case.
The inspection of the possible values of $Q$ allowed also to consider unlikely the value $M=(40\pm 5)M_{\odot}$ for \hde\ recently inferred from evolutionary calculation because it would allow  for high values of $i$ compatible with eclipses which, instead, are so far absent.
Thus, by reasonably assuming that
the Keplerian  period is an adequate model of the orbital period  we determined $i=(48\pm 7)$ deg. In this case, the relative semimajor axis  amounts to  $a=(42\pm 9)R_{\odot}$ ($\approx 0.2$ AU).

\section*{Appendix: the correction to the orbital period due to the quadrupole mass moment}
One of the six Keplerian orbital elements in terms of which it is
possible to parameterize the orbital motion in a
binary system is the mean anomaly  $\mathcal{M}$ defined as
$\mathcal{M}\equiv n(t-T_0)$, where $n$ is the mean motion and
$T_0$ is the time of pericenter passage. The mean motion $n\equiv
2\pi/ P_{\rm b}$ is inversely proportional to the time elapsed
between two consecutive crossings of the pericenter, i.e. the
anomalistic period $P_{\rm b}$. In Newtonian mechanics, for two
point-like bodies, $n$ reduces to the usual Keplerian expression
$n^{\rm Kep}=\sqrt{G{\rm M}/a^3}$, where $a$ is the semi-major axis of the
relative orbit  and
${\rm M}\equiv m_1 + m_2$ is the sum of the masses. In
many binary systems the period $P_{\rm b }$ is  accurately
determined in a phenomenological, model-independent way, so that
it accounts for all the dynamical features of the system, not only
those coming from the Newtonian point-like terms, within the
measurement precision.

Here we wish to calculate the contribution of the quadrupole mass moment $Q$ to the orbital period
in a general way.

By assuming that the spins and the orbital angular momentum are aligned, as done in \citep{Her95} for the \hde/\cyg\ system, the radial component of the acceleration induced by $Q$ can be cast into the form
\eqi
A_Q= \rp{3}{2}\rp{GQ}{r^4},\lb{radialacc}
\eqf
while the other component, i.e. the latitudinal one, vanishes.
$A_Q$ is small with respect to the
usual Newtonian monopole, so that it can be treated perturbatively. In
order to derive its impact on the orbital period $P_{\rm b}$, let
us consider the Gauss equation for the variation of the mean
anomaly in the case of an entirely radial disturbing acceleration
$A$ \eqi\dert{\mathcal{M}} t=n-\rp{2}{na}A
\left(\rp{r}{a}\right)+\rp{(1-e^2)}{nae}A\cos f,\lb{gauss}\eqf
where $f$ is the true anomaly, reckoned from the pericenter. After
inserting $A_Q$ into the right-hand-side of \rfr{gauss}, it must
be evaluated onto the unperturbed Keplerian ellipse \eqi
r=\rp{a(1-e^2)}{1+e\cos f}.\eqf By using \citep{Roy05}
\eqi
\dert
f{\mathcal{M}} =\left(\rp{a}{r}\right)^2\sqrt{1-e^2},\eqf
\rfr{gauss} yields
\begin{eqnarray}
\dert f t &=&\rp{n(1+e\cos f
)^2}{(1-e^2)^{3/2}} \left\{1+\rp{3GQ(1+e\cos f)^3}{n^2
a^5(1-e^2)^3}\right.\times\nonumber\\
& &\times\left.\left[\rp{\cos f(1+e\cos
f)}{2e}-1\right]\right\}.\lb{grossa}
\end{eqnarray}
%
%
The orbital period can
be obtained as
\begin{eqnarray} P_{\rm b}& \approx &
\rp{(1-e^2)^{3/2}}{n}\int_0^{2\pi}\left\{1-\rp{3GQ(1+e\cos
f)^3}{n^2 a^5(1-e^2)^3}\right.\times\nonumber\\
& &\times\left.\left[\rp{\cos f(1+e\cos
f)}{2e}-1\right]\right\}\rp{df}{(1+e\cos f)^2}.\lb{appro}
\end{eqnarray}
%
%
From \rfr{appro} it follows \eqi P_{\rm
b}\equiv P^{\rm Kep}+P^{Q},\eqf with
\begin{equation}\left\{\begin{array}{lll}
P^{\rm Kep}=2\pi\sqrt{\rp{a^3}{G{\rm M}}},\\\\
P^{Q}=\rp{3\pi Q}{\sqrt{Ga{\rm M}^3(1-e^2)^3}}.\lb{periodoQ}
\end{array}\right.\end{equation}
Note that $P^Q$ in \rfr{periodoQ} agrees with the expression of the anomalistic period of a satellite orbiting an oblate planet obtained by \citet{Cap05}: for a direct comparison $Q=-{\rm M}R^2 J_2$, where $J_2$ is the first even zonal harmonic of the multipolar expansion of the Newtonian part of the gravitational potential of the central body of mass M and equatorial radius $R$.

Solving for $Q$, we finally get \eqi Q=\rp{ P_{\rm b}\sqrt{Ga{\rm M}^3(1-e^2)^3}
}{3\pi}-\rp{2}{3}{\rm M}a^2(1-e^2)^{3/2}.\lb{QU}\eqf


\end{document}